\documentclass[aps,prb,twocolumn,superscriptaddress,longbibliography]{revtex4-2}
\usepackage[colorlinks=true,linkcolor=blue,citecolor=blue]{hyperref}
\usepackage{amsfonts}
\usepackage{subfigure}
\usepackage{amsmath}
\usepackage{txfonts}
\usepackage{amssymb}
\usepackage{amsbsy}
\usepackage{epsfig}
\usepackage{graphicx}
\usepackage{epstopdf}
\usepackage{mathdots}
\usepackage{color}
\usepackage{cleveref}
\usepackage{bbm}
\usepackage{natbib}
\usepackage{url}

\newcommand{\be}{\begin{equation}}
\newcommand{\ee}{\end{equation}}

\newcommand{\bs}{\boldsymbol}

\begin{document}

\title{Floquet topological phases with high Chern numbers in a periodically driven extended Su-Schrieffer-Heeger model} 
\author{Aayushi Agrawal}
\affiliation{Department of Physics, Birla Institute of Technology and Science, Pilani 333031, India}
\email{p20170415@pilani.bits-pilani.ac.in}
\author{Jayendra N. Bandyopadhyay}
\email{jnbandyo@gmail.com}
\affiliation{Department of Physics, Birla Institute of Technology and Science, Pilani 333031, India}

\date{\today}

\begin{abstract}

The high Chern number phases with the Chern number $|C| > 1$ are observed in this study of a periodically driven extended Su-Schrieffer-Heeger (E-SSH) model with a cyclic parameter. Besides the standard intra-dimer and the nearest-neighbor (NN) inter-dimer hopping of the SSH model, an additional next-nearest-neighbor (NNN) hopping is considered in the E-SSH model. The cyclic parameter, which plays the role of a synthetic dimension, is invoked as a modulation of the hopping strengths. A rigorous analysis of different phase diagrams has shown multiple Floquet topological phase transitions among the high Chern number phases. These phase transitions can be controlled by the strength and frequency of the periodic driving. Instead of applying perturbation theory, the whole analysis is done by Floquet replica technique. This gives a freedom to study high as well as low-frequency effects on the system by considering less or more number of photon sectors. This system can be experimentally realized through a pulse sequence scheme in the optical lattice setup.

\end{abstract}

\maketitle

\section{\label{sec:Introduction}Introduction}

The discovery of the integer quantum Hall (IQH) effect under a strong magnetic field \cite{klitzing1980new} unveiled a new family of materials called topological insulators (TIs) \cite{ryu2002topological, moore2010birth, hasan2010colloquium}. The TIs are mostly observed in compounds made of heavy elements with strong spin-orbit coupling. Unlike in the usual insulators, the topological insulators have conducting edges or surfaces with the insulating bulk part. The topological materials are classified based on the dimension and symmetries of the systems \cite{ryu2010topological, altland1997nonstandard, schnyder2008classification}. The topology of a system is quantified by topological invariant, which is a constant integer and does not vary under any continuous deformation of the energy bands (or Hamiltonian) \cite{hatsugai1993chern}. Besides some natural solid-state materials \cite{TopoMat01,TopoMat02}, the topological properties are experimentally simulated in ultra-cold atoms \cite{jotzu2014experimental, goldman2014light}, photonic systems \cite{rechtsman2013photonic}, etc. 

In a two-dimensional (2D) honeycomb lattice, Haldane introduced a complex next-nearest-neighbor (NNN) hopping to simulate a magnetic field-like effect. This leads to the realization of the quantum anomalous Hall (QAH) effect in the system \cite{haldane1988}. In this 2D system, a topological phase transition from a normal insulator to a Chern insulator is observed by varying the hopping phase and the sub-lattice potential difference \cite{haldane1988,haldane-related01,haldane-related02}. Recent investigations of realizing the Haldane model-like effect in lower-dimensional systems led to the so-called {\it extended} Su-Schrieffer-Heeger (E-SSH) model, which is described in one-dimension (1D). The original SSH model is the simplest possible 1D system which shows topological property \cite{ryu2010topological,asboth2016short}. This model was proposed to describe the electronic properties of the poly-acetylene chain \cite{su1979solitons, heeger1988solitons}. Following the idea of Haldane, in the E-SSH model, in addition to a standard nearest-neighbor (NN) hopping, the NNN hopping term is introduced \cite{li2014topological}. Furthermore, the hopping strengths are modulated by a cyclic variable that plays the role of an additional synthetic dimension. This synthetic dimension, together with the momentum, defines an effective 2D space on a real 1D space \cite{kraus2012topological,verbin2013observation,mei2014topological}. Consequently, instead of the winding number, one has to now calculate the Chern number as a topological invariant for this effective 2D system \cite{thouless1983quantization}. In the E-SSH model, if we also introduce an onsite potential, then one can map the synthetic dimension, and the momentum of this exactly with two momenta of the Haldane model \cite{li2014topological}. Therefore, this mapping gives a clear understanding of the relation between the E-SSH model and the Haldane model. However, instead of two modulated terms (the onsite and the NNN) together with the SSH Hamiltonian, it is experimentally easier to realize the version of the E-SSH model that has only the modulated NNN term with the SSH Hamiltonian. Recently, some other versions of the E-SSH models are also investigated \cite{guo2015kaleidoscope, bahari2016zeeman, ahmadi2020topological, li2019extended}.
 
The QAH effect is experimentally realized mostly in magnetic topological insulators \cite{MTI01,MTI02,MTI03,MTI04,MTI05,MTI06} and in bilayer graphene twisted with a magic angle \cite{BLG01,BLG02}. In all these investigations, the QAH is observed only with the Chern number $C = \pm 1$. Very recently, the QAH effect is observed with a tunable Chern number up to $C = 5$ \cite{QAH-HighChern}. In this work, a system of multiple layers with alternating magnetic and undoped topological insulators was fabricated to observe high Chern numbers. Interestingly, the number of undoped topological layers decides the Chern number of the system.

Recent studies have established that the nontrivial topological properties can be ascribed to some topologically trivial systems by applying external time-periodic driving \cite{cayssol2013floquet, kitagawa2010topological, gomez2013floquet, rudner2019floquet}. Since the periodically driven systems are studied under Floquet theoretical framework, these synthetic topological systems are called {\it Floquet topological} systems. Many times, the Floquet systems show new exotic topological phases that may not be realized by any static means. For example, new Floquet topological phases in graphene illuminated by circularly polarized laser field \cite{usaj2014irradiated, kundu2014effective}, higher-order topological phases in superconductor \cite{FHOTI-02, FHOTI-03}, switching of the native topology of the SSH model \cite{dal2015floquet}, etc.

Very recently, it has been observed that periodic driving can generate topological insulators with larger Chern numbers in higher dimensional (higher than 1D) systems \cite{FHChern01,FHChern02,FHChern03,FHChern04,FHChern05,FHChern06,FHChern07,
FHChern08,FHChern09,FHChern10}. Since the 1D E-SSH model can be mapped to the 2D Haldane model, a natural question is whether one can realize larger Chern numbers in the Floquet E-SSH model. The richer topological phases with higher Chern numbers are obtained when the SSH system is periodically modulated in time \cite{lu2019topological}. Moreover, the topological phases with larger Chern numbers were reported in a Dirac delta-kicked SSH model with onsite potential \cite{li2020topological}. Here, we have applied smooth sinusoidal drivings to the E-SSH model having the usual NN hopping of the SSH model and an additional NNN hopping to realize the topological phases with large Chern number.  

We have planned this paper in the following way: In Sec. \ref{sec:Model}, we discuss the undriven E-SSH model and introduce the Floquet version of the model with sinusoidal periodic driving. This section also discusses theoretical background like the underlying symmetries in the system and its consequence in the topology of the system. In the next section, Sec. \ref{sec:Results}, we present all the results. We conclude with a final remark in Sec. \ref{sec:Final remarks}.  

\section{\label{sec:Model}Model and theoretical background}

This section starts with the description of the E-SSH model. In this paper, we consider the same version of the {\it static} E-SSH model, which was originally proposed in Ref. \cite{li2014topological}. Here we discuss the salient features of this model, and then introduce the periodically driven version of this. We also present a basic theoretical background that is needed to quantify and analyze the results.

\subsection{Undriven E-SSH model}

\begin{figure}[t]
\centering
\includegraphics[width=8.8cm,height=5cm]{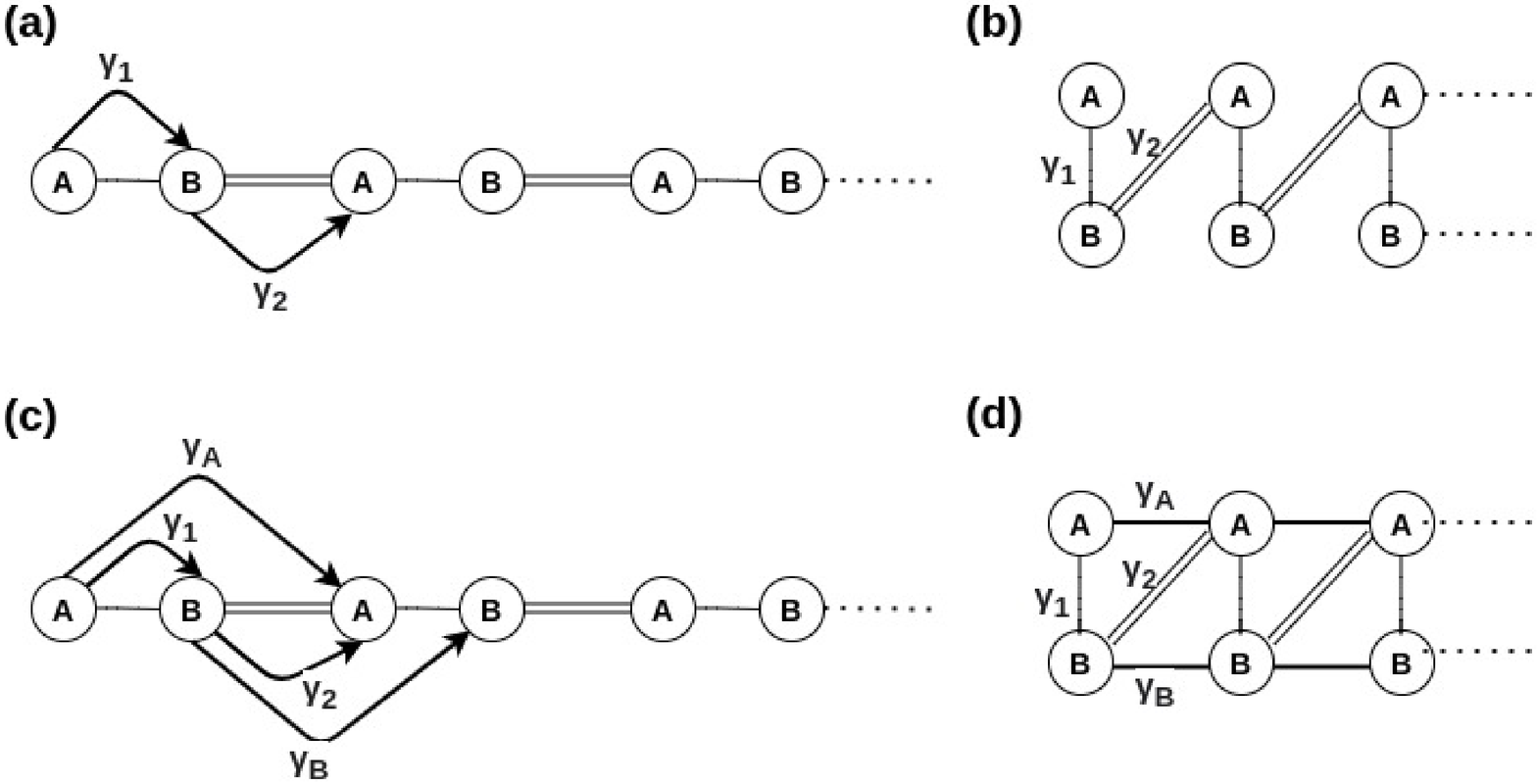}
\caption{\label{Model-Fig} The original SSH model can be described by two geometrical structures: (a) a linear chain and (b) a ladder geometry. The same is shown for the E-SSH model in (c) and (d).}
\end{figure}

The original SSH model is a 1D chain that consists of two sub-lattices, represented by sites $A$ and $B$ in every unit cell. In this model, as shown in Fig. \ref{Model-Fig}(a), only intra-cell hopping among $A$ and $B$ sites and the NN inter-cell hopping are allowed. Figure \ref{Model-Fig}(b) shows that the same dynamics can also be described on a ladder geometry by placing each dimer vertically. For the E-SSH model, an additional NNN hopping from the $A$-site ($B$-site) of a unit cell to the $A$-site ($B$-site) of the neighboring unit cell is included. The E-SSH model can also be described as a 1D chain and a ladder. These are shown respectively in Figs. \ref{Model-Fig}(c) and \ref{Model-Fig}(d). Following \cite{li2014topological}, we consider the Hamiltonian of the undriven E-SSH model as
\begin{equation}
\begin{split}
H_{E-SSH} &= \underbrace{-\sum_n \big\{\gamma_1 c_{A,n}^{\dagger} c_{B,n} + \gamma_2 c_{A,n+1}^{\dagger} c_{B,n}\big\} + {\rm h.c}}_{\equiv\, H_{SSH}}\\
&+ \underbrace{\sum_n \big\{\gamma_A c_{A,n+1}^{\dagger} c_{A,n} + \gamma_B c_{B,n+1}^{\dagger} c_{B,n}\big\} + {\rm h.c.}}_{\equiv\, H_{NNN}},
\end{split}
\label{Ex_SSH_Ham}
\end{equation}
where $H_{SSH}$ is the standard SSH Hamiltonian and $H_{NNN}$ is the NNN hopping part of the Hamiltonian. The operators $c_{A,n}$ ($c_{B,n}$) and $c_{A,n}^{\dagger}$ ($c_{B,n}^{\dagger}$) are the annihilation and creation operators defined on the sub-lattice $A (B)$. Here the parameters 
\[\gamma_1 = (1+\delta \cos \theta)~~ {\rm and} ~~\gamma_2 = (1-\delta \cos \theta)\] 
respectively denote the intra-cell and inter-cell hopping strengths. Here the parameter $\delta$ is always set as positive. In the SSH model, the relative strengths of the hopping amplitudes decide the topological property of the system. Here we have modulated the hopping amplitudes by a single cyclic parameter $\theta$. Therefore, the Hamiltonian becomes periodic in $\theta$, i.e., $H_{E-SSH}(\theta) = H_{E-SSH}(\theta+2\pi)$. The cyclic parameter $\theta$ can be considered as an additional synthetic dimension. 

In the NNN hopping part, the following form of the hopping strengths are considered: \[\gamma_A = g_A+h \cos (\theta + \phi) ~~{\rm and}~~ \gamma_B = g_B+h \cos (\theta - \phi),\] where the constant parameters $(g_A, g_B)$ and the modulation strength $h$ are introduced to avoid the overlap between the two bands. Moreover, a parameter $\phi$ is introduced to observe a topological phase transition at $\phi=0$. This phase transition is identified by the change of sign of the Chern number at this point.  

In order to observe the properties of the bulk energy of the E-SSH model, periodic boundary condition (PBC) is imposed on the system. The PBC allows the transformation of the Hamiltonian from the real space to the momentum space ($k$-space) via Fourier transformation. 
\begin{equation}
c_{A/B,n} = \frac{1}{\sqrt{N}} \sum_k e^{ikn}\, \tilde{c}_{A/B,k}, 
\end{equation}
where $N$ is the number of sites in each sub-lattice and this number is also equal to the number of dimers. Substituting these Fourier transformed operators in Eq. \eqref{Ex_SSH_Ham} and defining Nambu spinors $\Psi_k = \left(\tilde{c}_{A,k} ~~ \tilde{c}_{B,k}\right)^T$, we write down the Hamiltonian in the $k$-space as
\begin{subequations}
\begin{equation}
\begin{split}
&H_{\rm E-SSH} = \sum_k \Psi_{k}^{\dagger}\, \mathcal{H}_{\rm E-SSH}(k) \, \Psi_{k},\\{\rm with}~ &\mathcal{H}_{\rm E-SSH}(k) = h_{k0} \mathbbm{1} + \bs{h}(k) \cdot \bs{\sigma},
\label{eq:E-SSH-Ham01}
\end{split}
\ee
where $\mathcal{H}_{\rm E-SSH}(k)$ is the Bloch Hamiltonian and $\bs{\sigma} = \left(\sigma_x,~\sigma_y,~\sigma_z\right)$ are Pauli (pseudo) spin operators defined in the sub-lattice degrees of freedom. The parameter $h_{k0}$ and the components of the parameter vector $\bs{h}(k) = \left(h_{kx},~h_{ky},~h_{kz}\right)$ are given as
\be
\begin{split}
h_{k0} &= (\gamma_A + \gamma_B) \cos k,~h_{kx} = -(\gamma_1 + \gamma_2 \cos k) \\ h_{ky}&= -\gamma_2 \sin k, ~ h_{kz} = (\gamma_A - \gamma_B)\cos k.
\end{split}
\label{eq:E-SSH-Ham02}
\ee
\label{eq:E-SSH-Ham}
\end{subequations}

\subsubsection{\label{sec:Symmetries} Symmetries in the system}

The SSH Bloch Hamiltonian $H_{SSH}(k) \equiv \bigl.\mathcal{H}_{\rm E-SSH}(k)\bigr|_{\gamma_A=\gamma_B=0}$ follows three fundamental symmetries: 
\begin{subequations}
\begin{align}
\mathcal{T}^{-1}\, \mathcal{H}_{\rm SSH}(k)\, \mathcal{T} &= \mathcal{H}_{\rm SSH}(-k), ~~~~{\rm (Time-reversal)}\\
\mathcal{P}^{-1}\, \mathcal{H}_{\rm SSH}(k)\, \mathcal{P} &= -\mathcal{H}_{\rm SSH}(-k), ~~~~{\rm (Particle-Hole)}\\
\mathcal{C}^{-1}\, \mathcal{H}_{\rm SSH}(k) \, \mathcal{C} &= -\mathcal{H}_{\rm SSH}(k), ~~~~{\rm (Chiral)}.
\end{align}
\end{subequations}
Here, the time-reversal operator $\mathcal{T} = K$, the complex conjugation operator, and the particle-hole operator $\mathcal{P} = \sigma_z\, K$ are anti-unitary operators. The other symmetry operator, the chiral operator, $\mathcal{C} = \mathcal{P\, T} = \sigma_z$ is a unitary operator. Even though this is a unitary symmetry, but it does not follow the standard commutation relation $\bigl[\sigma_z, \,\mathcal{H}_{\rm SSH}(k)\bigr] = 0$, rather it follows the anti-commutation relation $\bigl\{\sigma_z, \,\mathcal{H}_{\rm SSH}(k)\bigr\} = \sigma_z \,\mathcal{H}_{\rm SSH}(k) + \mathcal{H}_{\rm SSH}(k) \,\sigma_z = 0$. The chiral symmetry appears in the SSH model due to the absence of any hopping within the sub-lattices. Consequently, the parameter vector $\bs{h}(k)$ of the SSH Hamiltonian lies on the $x-y$ plane; and depending on whether the tangent of $\bs{h}(k)$ encloses the origin with the variation of $k$, the SSH model shows its topological property. This topological property is quantified by the winding number, which counts how many times the tangent of $\bs{h}(k)$ encircles the origin as we vary $k$ within the Brillouin zone (BZ), i.e., $k=-\pi$ to $\pi$. Since the SSH Hamiltonian satisfies these three symmetries, this system is classified as the BDI class of the {\it ten-fold way of classification} of the quantum topological matters \cite{ryu2010topological, RMP-TopoClassify, bandyopadhyay2019temporal}. 

The addition of the NNN term $H_{NNN}$ with the Hamiltonian $H_{SSH}$ introduces a hopping within the sub-lattices. Consequently, this additional term breaks the chiral symmetry in the E-SSH model. Unlike the SSH case, due to the presence of $\sigma_z$ in the Bloch Hamiltonian of the E-SSH model, the parameter vector $\bs{h}(k)$ does not lie on the $x-y$ plane, and therefore the tip of this vector cannot enclose the origin. Consequently, the winding number becomes ill-defined for this system. However, in this model, the modulation parameter $\theta$ can be varied adiabatically from $0$ to $2\pi$ without closing the gap between the bands for any non-zero value of the parameter $\phi$. Since, the modulation parameter $\theta$ plays the role of a synthetic dimension, the E-SSH model becomes an effective 2D system defined on the periodic $(k,\,\theta)$-space, which is geometrically identical to a torus. The topological property of a system higher than 1D is generally quantified by topological invariants known as Chern numbers. The {\it first} Chern number or just the Chern number $C$ is calculated by integrating the Berry curvature over the BZ, which is a closed 2D manifold, as
\begin{equation}
C = \frac{1}{2\pi} \int_{BZ} \mathcal{F}(k,\theta)\, dk\, d\theta,
\label{eq:ChernNumber}
\end{equation}
where the Berry curvature $\mathcal{F}(k,\theta) = (\partial_{\theta} A_k - \partial_k A_{\theta})$. The Berry curvature is defined via Berry connections $(A_k,\,A_\theta)$, which are defined as \cite{berry1984quantal}
\[A_\eta \equiv \langle \psi (k,\theta)|\,\partial_\eta\,| \psi(k,\theta) \rangle, ~{\rm where}~ \eta = k, \theta ;\] and $|\psi(k,\theta) \rangle$ are the Bloch functions. 

Depending upon the parameters $\gamma_A$ and $\gamma_B$, symmetries of the E-SSH model changes \cite{li2014topological}. We can see from the E-SSH Hamiltonian in Eqs. \eqref{eq:E-SSH-Ham}, when $\gamma_A = \gamma_B$, the $\sigma_z$ term vanishes in the Bloch Hamiltonian. For this case, in addition to the chiral symmetry $\mathcal{C}$, the particle-hole symmetry $\mathcal{P}$ also breaks down. This symmetry breaking results in an asymmetry in the energy bands and appearance of non-zero degenerate edge states. For $\gamma_A = -\gamma_B$ case, instead of $\sigma_z$ term, now the term with the identity matrix $\mathbbm{1}$ vanishes. Therefore, instead of the particle-hole symmetry, now the time-reversal symmetry $\mathcal{T}$ breaks down in the system. In this case, the energy spectrum will again be symmetric, but now the degeneracy of the non-zero edge states will be lifted. Finally, for the general case, when the parameters $\gamma_A$ and $\gamma_B$ do not satisfy any special relation, then the three symmetries mentioned above break down in the E-SSH model. For this case, the energy bands will be asymmetric, and the edge states will be non-zero and non-degenerate. In this paper, we consider the general case $|\gamma_A| \neq |\gamma_B|$, where none of the symmetries is preserved.

\subsection{\label{sec:Driven-E-SSH} Periodically driven E-SSH model}

We now describe the periodically driven E-SSH model which we extensively study as a prototype model for a realization of the high Chern number topological phases. The Hamiltonian of this model is 
\begin{subequations}
\begin{equation}
\hat{H}_{E-SSH}(t) = \hat{H}_{E-SSH} + \hat{V}(t),
\end{equation}
where $\hat{H}_{E-SSH}$ is already defined in Eq. \eqref{eq:E-SSH-Ham} and the time-periodic potential satisfies the condition $\hat{V}(t+T) = \hat{V}(t)$. Consequently, the total Hamiltonian is also a time-periodic $\hat{H}_{E-SSH}(t+T) = \hat{H}_{E-SSH}(t)$. The time-periodic driving part of the Hamiltonian is considered as 
\begin{equation}
\begin{split}
\hat{V}(t) &= \mathit{v}_1(t) \sum_n (c_{A,n}^{\dagger} c_{B,n} - c_{A,n+1}^{\dagger} c_{B,n})\\
 &+ \mathit{v}_2(t) \sum_n (c_{A,n+1}^{\dagger} c_{A,n} - c_{B,n+1}^{\dagger} c_{B,n}) + h.c.
\end{split}
\end{equation}
\end{subequations}
Here, the periodic driving: $v_1(t) = 2 V_1 \cos \omega t$ and $v_2(t) = 2 V_2 \cos (\omega t + \alpha)$, where $V_1$ and $V_2$ are the driving amplitudes, the frequency of both the driving is $\omega$, and $\alpha$ is a phase factor. In the momentum space, the Bloch Hamiltonian of the time-periodic E-SSH model will be of the form
\begin{widetext}
\begin{equation}
\mathcal{H}_k(t) = (\gamma_{A}+\gamma_{B})\cos k\, \mathbbm{1} - \left[(\gamma_{1} + \gamma_{2} \cos k) - \mathit{v}_1(t)(1-\cos k)\right] \sigma_x \,-\,
 \left[ \gamma_{2} + \mathit{v}_1(t)\right] \sin k\, \sigma_y \,+\, \left[\gamma_{A} - \gamma_{B} + 2 \mathit{v}_2(t) \right]\, \cos k\, \sigma_z.
\end{equation}
\end{widetext}
The driving protocol is chosen in such a way that it introduces the time-dependent functions only with the pseudo-spin operators $\bs{\sigma}$. 

\subsubsection{\label{sec:Floquet bands}Floquet analysis of the driven E-SSH model}

According to the ansatz of Floquet theory, the solution of the time-dependent Schr\"odinger equation (TDSE) for the periodically driven E-SSH Hamiltonian 
\begin{subequations}
\be
i \frac{d}{dt}\, |\psi(t) \rangle = H_{E-SSH}(t)\, |\psi(t)\rangle
\label{eq:TDSE}
\ee
will be of the form 
\be
|\psi_n(t)\rangle = e^{-i \epsilon_n t}\,|u_n(t)\rangle,~{\rm where}~ |u_n(t)\rangle = |u_n(t+T)\rangle.
\ee
\end{subequations}
Here we set $\hbar=1$ and $T = 2\pi/\omega$ is the time-period. The states $|\psi_n(t)\rangle$ are known as Floquet states \cite{shirley1965solution}. The time-periodic states $|u_n(t)\rangle$ are called Floquet modes and $\epsilon_n$ represents the quasi-energy of the $n$-th state. The Floquet states are temporal analogue of the Bloch states of the solid state physics \cite{Kittel2004,Ashcroft76}, where the Hamiltonian has the periodicity in space. 
  
The Floquet states are the eigenstates of the time-evolution operator over one driving period, i.e., \[U(t_0+T,\, t_0)\, |\psi_n(t_0)\rangle = e^{-i \epsilon_n T}\, |\psi_n(t_0)\rangle,\] where $t_0$ is an arbitrary initial time. Therefore, one can obtain the Floquet states and the quasienergies by diagonalizing $U(t_0+T,\, t_0)$. The eigenvalues $e^{-i \epsilon_n T}$ and the corresponding eigenstates $|\psi_n(t_0)\rangle$ are gauge independent, but the quasienergy $\epsilon_n$ and the corresponding Floquet mode $|u_n(t)\rangle = e^{i \epsilon_n t}\,|\psi_n(t)\rangle$ are dependent on the gauge. By introducing an integer label $m = \pm 1,\, \pm 2, \dots$, we get all the possible gauges of the quasienergies $\epsilon_{nm} = \epsilon_n + m \omega$ and the corresponding Floquet modes $|u_{nm}(t)\rangle = e^{i m \omega t}\,|u_n(t)\rangle$, which give the same Floquet state $|\psi_n(t)\rangle$. Substituting, the general representation 
\be
|\psi_n(t)\rangle = e^{-i \epsilon_{nm} t}\,|u_{nm}(t)\rangle
\ee 
in the TDSE [Eq. \eqref{eq:TDSE}], we get the Floquet Hamiltonian
\be
\mathcal{H}_F:\, \left[i \frac{d}{dt} - H_{E-SSH}(t)\right]\, |u_{nm}(t)\rangle = \epsilon_{nm}\,|u_{nm}(t)\rangle.
\ee
The above relation forms an eigenvalue problem in an extended Hilbert space $\mathbb{H} \otimes \mathbb{T}$. The extended Hilbert space, also known as the Sambe's space, where $\mathbb{H}$ is the usual Hilbert space and $\mathbb{T}$ is a Hilbert space spanned by the time-periodic functions $e^{-im\omega t}$ where $m \in \mathbb{Z}$ \cite{Sambe}. The time-periodic property of the Hamiltonian $H_{E-SSH}(t)$ and the Floquet modes $|u_n(t)\rangle$ allow us to expand these in the Fourier series. This process splits the driven spectrum into an infinite number of copies of undriven Hamiltonian separated by $m \omega$. This process is called the Floquet replica method, which is analogous to the dressed atom picture of the electromagnetic field induced atomic systems \cite{eckardt2008dressed}. Therefore $m$ is called the photon numbers. In the Sambe space, the general representation of the Floquet Hamiltonian is:\begin{subequations}
\be
\mathcal{H}_F = 
\begin{bmatrix}
\ddots & \vdots& \vdots & \vdots&\vdots &\vdots &\iddots  \\
\dots& H_0 - 2\omega  & H_{-1}& H_{-2}& H_{-3} &H_{-4}&\dots\\
\dots&H_1 & H_0 - \omega & H_{-1} & H_{-2} & H_{-3} &\dots\\
\dots&H_2 & H_1 & H_0 & H_{-1} & H_{-2} & \dots\\
\dots& H_3& H_2 & H_1 & H_0 +  \omega &H_{-1}&\dots\\
\dots&H_4 &H_3 &H_2 & H_1& H_0 + 2 \omega &\dots\\
\iddots& \vdots&\vdots &\vdots &\vdots &\vdots & \ddots
\end{bmatrix},
\ee
where for the E-SSH model 
\be
H_m = \frac{1}{T} \int_0^T H_{E-SSH}(t)\,e^{-im\omega t}\,dt.
\label{Hm}
\ee 
\end{subequations}
By construction/definition, the Floquet Hamiltonian is exact if one considers all infinite number of photon sectors. However, we have to consider a finite number of photon sectors in numerics. The number of photon sectors is decided by the strength of the driving frequency $\omega$. Suppose the driving frequency is of the order or smaller than the characteristic frequencies of the undriven system, one has to consider a larger number of photon sectors to get convergence in the calculation of the (dressed) Floquet bands around {\it zero} photon sector. On the other hand, one has to consider only a few photon sectors for a very high frequency. In this paper, we need to consider maximum {\it eleven} photon sectors to achieve numerical convergence. Since, our driving protocol is monochromatic (absence of any harmonics of $\omega$), $H_m = 0$ for $|m| \ge 2$ and for the remaining values $m=0,\,\pm1$ we have in the $k$-space
\begin{equation}
\begin{aligned}
H_{0,k} &= (\gamma_{A} + \gamma_{B}) \cos k \ \mathbbm{1} -(\gamma_{1} + \gamma_{2} \cos k)\, \sigma_x\\ &- \gamma_{2} \sin k\, \sigma_y
+ (\gamma_{A} - \gamma_{B}) \cos k\, \sigma_z, \\
H_{\pm1,k} &=  V_1  \left[(1-\cos k)\, \sigma_x - \sin k\, \sigma_y \right]\\ &+ 2 V_2\, e^{\pm i\alpha} \cos k \,  \sigma_z.
\end{aligned}
\end{equation}
Here, the Hamiltonian $H_{0,k}$, defined in $m=0$ photon sector, is just the undriven E-SSH Hamiltonian represented in the $k$-space. In Sambe's space representation, the periodic driving introduces the off-diagonal blocks $H_{\pm1}$ in the Floquet Hamiltonian, which connects different photon sectors. In the $k$-space, the eigenstates of $H_F$ form Floquet-Bloch bands where both (quasi)energy and momentum are periodic. In the crystalline solid, the Bloch bands hybridize and develop bandgaps at the crossing points \cite{Kittel2004,Ashcroft76}. Similarly, the crossing points between different photon sectors of the Floquet-Bloch bands open dynamical gaps \cite{Sambe}. The Floquet-Bloch bands are also experimentally observed on the surface of a 3D topological insulator Bi$_2$Se$_3$ \cite{Floquet-Bands}. Again the Chern number can be the quantifier of the topology of the Floquet-Bloch bands. For that, in Eq. \eqref{eq:ChernNumber}, one has to replace the Bloch bands by the Floquet-Bloch bands.

\section{\label{sec:Results} Results}

In this section, we present all the results. First, we discuss the Floquet (energy) band diagram of the E-SSH model and study their topological properties by calculating the Chern number. We then discuss how the Floquet topological phase transition in the system is dependent on the parameters of the undriven system. Later, we discuss the role of the driving frequency and the amplitude on the Floquet topological phase transition. Finally, we concentrate on the parameter regimes where the topological phases with higher Chern numbers $C=3$ and $C=5$ are observed. 

\subsection{Floquet (energy) band diagrams of the E-SSH model and its topological property}

\begin{figure}[t]
\centering
\includegraphics[width=8.8cm,height=7cm]{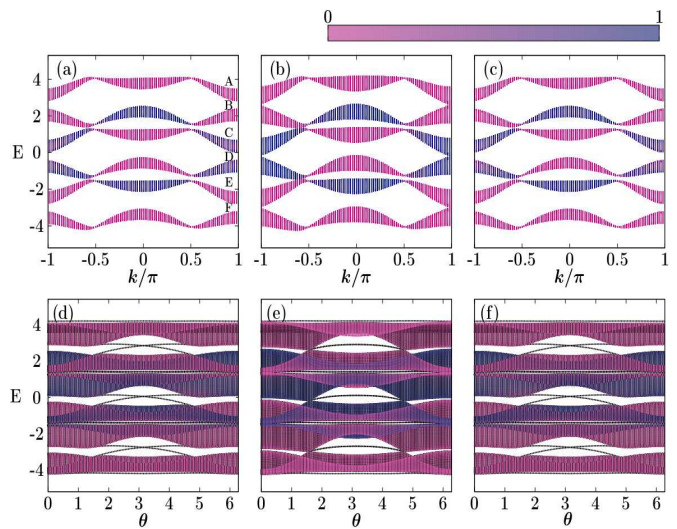}
\caption{\label{Fig Floquet bands} Floquet bands of the driven extended SSH model at the high-frequency regime. The parameters $\delta = 0.3$, $\omega = 2.8$, and $V = 0.1$ are kept fix for all the cases. In the high-frequency range, we use $5$ photon sectors in the numerics and present the Floquet bands corresponding to the central three photon sectors with $m = 0, \pm1$. The color bar represents the static weight ($|\langle u_{\alpha}|u_{\alpha}\rangle|^2$) and separates the zero photon sector from the higher photon sectors. In the upper panel, $E$ is plotted as a function of $k$; while in the lower panel, $E$ is plotted against $\theta$. We set $\phi = -\pi/4$ for the left side figures, $\phi = 0$ for middle figures, and $\phi = \pi/4$ for the right side figures. Nonzero value of $\phi$ induces a additional gap at $k = \pm \pi$ which is absent for $\phi = 0$. Varying $\phi$, topological phase transitions are exhibited via the closing and opening of the Floquet bands. The lower panel shows the quasienergies for the open boundary condition and the edge states presence.}
\end{figure}

The Floquet band diagram for the driven E-SSH model with driving frequency $\omega=2.8$ are presented in Fig. \ref{Fig Floquet bands}(a)-(c) respectively for $\phi = -\pi/4$, $0$ and $\pi/4$ as a function of momenta $k$. For the each case, the Floquet-Bloch bands are presented for three ($m=0,\,\pm1$) photon sectors. Here the synthetic dimension $\theta \in [0,\,2\pi]$ plays the role of a parameter. Since the driving frequency is much larger than the maximum bandgap of the undriven system, this value of $\omega$ can be considered as a high-frequency case. Therefore, we find that the calculation with only five photon sectors ($m=0,\,\pm1,\,\pm2$) is sufficient for the numerical convergence of the three central Floquet bands. In all the calculation, we set $V_1 = V_2$, the phase factor $\alpha = \pi/2$, and the dimerization constant $\delta = 0.3$. Hereafter, we replace all $V_1$ and $V_2$ by a single parameter $V$. For Fig. (\ref{Fig Floquet bands}), we set $V=0.1$ and the other parameters as $g_A = g_B = 0.1$ and $h = 0.2$. Here, we note that the bulk spectra for $\phi = \pi/4$ and $-\pi/4$ are identical. We also notice a pair of gaps between the Floquet bands at $k = \pm \pi/2$ and another pair of gaps at the BZ boundary $k = \pm \pi$. The gaps at the BZ boundary disappear at $\phi = 0$. This opening and closing of the Floquet bandgaps indicate a topological phase transition in the system. 

In Fig. \ref{Fig Floquet bands}(d)-(f), we have presented the Floquet energy spectrum of the central three Floquet bands for the open boundary condition as a function of the cyclic parameter $\theta$ for the same values of $\phi$. These show the presence of edge states in the system, a typical property of any topologically nontrivial system. Here again, Figs. \ref{Fig Floquet bands}(d) and (f) show the identical nature of the energy spectrum, whereas Fig. \ref{Fig Floquet bands}(e) shows the closing off of the {\it bandgap} at $\theta = \pi/2$ and $3\pi/2$.

According to  Fig. \ref{Fig Floquet bands}(a) and (c) or  Fig. \ref{Fig Floquet bands}(d) and (f), the bulk spectrum for $\phi = -\pi/4$ and $\phi = \pi/4$ do not reveal different topological property of the Floquet bands. Therefore, we need to calculate the Chern number to detect the topological phase transition. We follow Ref. \cite{fukui2005chern} to evaluate the Chern number. In Fig. \ref{Fig Floquet bands}(a), where $\phi = -\pi/4$,
\begin{equation}
 C_{\beta} = \bigg\{
\begin{matrix}
&1 & (\beta = B,D,F)\\
&-1 & (\beta = A,C,E).
\end{matrix} 
\end{equation}
Here $C_{\beta}$ is the Chern number of the Floquet-Bloch band $\beta$, as shown in Fig. \ref{Fig Floquet bands}(a). Similarly, for $\phi = \pi/4$ [Fig. \ref{Fig Floquet bands}(c)], we get
\begin{equation}
 C_{\beta} = \bigg\{
\begin{matrix}
&1 & (\beta = A,C,E)\\
&-1 & (\beta = B,D,F)
\end{matrix} 
\end{equation}
We obtain the Chern number of opposite signs for the Floquet bands corresponding to two opposite values $\phi = \pi/4$ and $\phi = -\pi/4$. For any arbitrary nonzero values of $\phi$ with opposite sign, we always find $C=\pm 1$. This indicates topological phase transition at $\phi=0$. The same results were obtained even for the undriven E-SSH model \citep{li2014topological}.

In the remaining part of this section, we explore the role of periodic driving in the topological phase transition. We particularly show the advent of Floquet-Bloch bands with higher Chern numbers due to periodic driving.

\subsection{\label{Phase diagram I} Phase diagram between $g_{-}/h$ and $\phi$}

\begin{figure}[t]
\includegraphics[width=8.5cm,height=6.5cm]{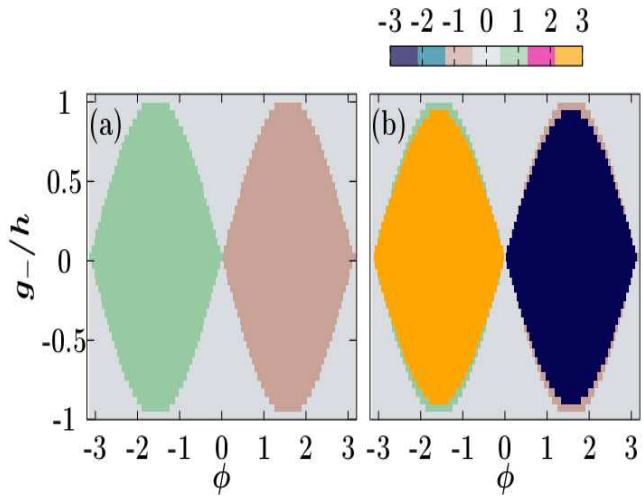}
\caption{\label{Fig Phase diagram I} Phase diagram for the topological phase transition is shown via plotting the sum of the Chern numbers $C_{\rm s}$ of all the Floquet bands below $E=0$ as a function of $g_{-}/h$ and $\phi$. Here, {\it nine} photon sectors are required to get numerical convergence. In (a) $V = 0.1$ and in (b) $V = 1.0$ are considered for a fixed $\omega = 2.8$. The color bar represents $C_{\rm s}$.}
\end{figure}

We now examine the effect of the NNN hopping and the time-periodic driving on the topological properties of the E-SSH model. In Fig. \ref{Fig Phase diagram I}, we present a phase diagram to show the topological phase transition in the driven E-SSH model. We define a parameter $g_- = (g_A - g_B)/2$, which gives a measure of the relative strength of two intra-sublattice hopping ($A$ to $A$ and $B$ to $B$). The parameter $g_-$ is then scaled with respect to the modulation parameter $h$, where we fix $h=0.2$. In Fig. \ref{Fig Phase diagram I}, we consider two different driving amplitudes $V = 0.1$ and $V = 1.0$ for a fixed frequency $\omega = 2.8$. Here we plotted the sum of the Chern numbers, denoted by $C_{\rm{s}}$, of all the Floquet bands below $E=0$ as a function of $g_{-}/h$ and $\phi$. Since here we are considering high-frequency driving, we find that the consideration of five photon sectors is sufficient for the convergence. Nevertheless, here we have presented the results with nine photon sectors.

In Fig \ref{Fig Phase diagram I}(a), we present the result for the driving amplitude $V = 0.1$. For this case, the phase diagram is showing three topologically distinct regions: a non-trivial region within $\phi \in [-\pi,0)$ with $C_{\rm s} = +1$; another non-trivial region within $\phi \in (0,\pi]$ with $C_{\rm s}=-1$, and the rest is topologically trivial region with $C_{\rm s}=0$. If we compare the phase diagram of the driven E-SSH model with the undriven E-SSH model, discussed in Ref. \cite{li2014topological}, then we observe that the periodic driving flips the sign of the Chern number. One should also note that the phase diagram of the E-SSH model shares qualitatively similar properties with the Haldane model \cite{haldane1988}. In Fig. \ref{Fig Phase diagram I}(b), we consider the same phase diagram corresponding to the stronger driving field $V=1.0$. For this case, we observe appearance of higher Chern number $C_{\rm s}=\pm 3$ in the same parameter regime where we observed $C_{\rm s} = \pm 1$ for the weaker driving strength $V=0.1$. Interestingly, here we notice that $C_{\rm s}=+3$ ($C_{\rm s}=-3$) region is surrounded by a thin layer of $C_{\rm s} = +1$ ($C_{\rm s}=-1$) region. Overall these phase diagrams suggest two types of topological phase transition depending on the periodic driving strengths: for the weaker driving, the Chern number flips its sign; and for the stronger field, the Chern number not only flips its sign but also increases the magnitude. 

\subsection{\label {phase diagram II} Topological phase transition from a non-trivial phase to another non-trivial phase}

In the previous subsection \ref{Phase diagram I}, we have shown two different types of topological phase transitions depending on the driving strengths at the high-frequency limit. We now explore the simultaneous role of the driving strength and the frequency on the topological phase transition in a more detail. More precisely, we study the variation of the sum of the Chern number $C_{\rm s}$ as a function of the driving strength $V$ and the frequency $\omega$. We have found that, by tuning the driving amplitude and frequency, one can get $C_{\rm s}$ of different magnitudes. 

Here we cover a wide range of frequencies: from a much smaller frequency than the lowest bandgap energy of the undriven E-SSH model to a frequency much higher than the largest bandgap of the same undriven system. We present the phase diagram for the lower frequency regime $(0.01 \leq \omega \leq 2.0)$ and the higher frequency range $(\omega > 2.0)$ separately. Here, we set the parameter $\phi = -\pi/4$ and keep it fixed throughout this analysis. Except at $\phi = 0$, we get a qualitatively similar topological property for any other values of $\phi$. We also notice that the Chern number changes sign with the sign change of $\phi$.

\begin{figure}[b]
\includegraphics[width=8.5cm,height=6.5cm]{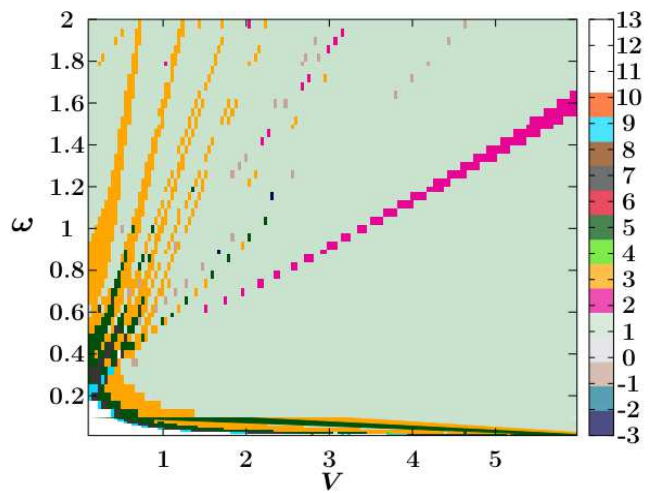}
\caption{\label{Fig Phase diagram II low-freq} The summed over Chern number $C_{\rm s}$ is plotted as a function of $\omega$ and $V$ in the low-frequency regime $\omega \leq 2$. Here {\it eleven} photon sectors are considered for the convergence. The other parameter are set at: $\phi = -\pi/4$, $g_A = g_B = 0.1$, and $h = 0.2$. The color bar represents the Chern number $C_{\rm s}$. For very small values of $\omega$ and $V$, the sum of the Chern number of the Floquet bands is very high which is shown with the white region. A major part of the phase diagram is occupied by the lower Chern number phases with $C_{\rm s} = + 1$. Topological phases with high Chern numbers $|C| > 1$ are demonstrated by different color coding.}
\end{figure}

\begin{figure}[t]
\includegraphics[width=8.5cm,height=6.5cm]{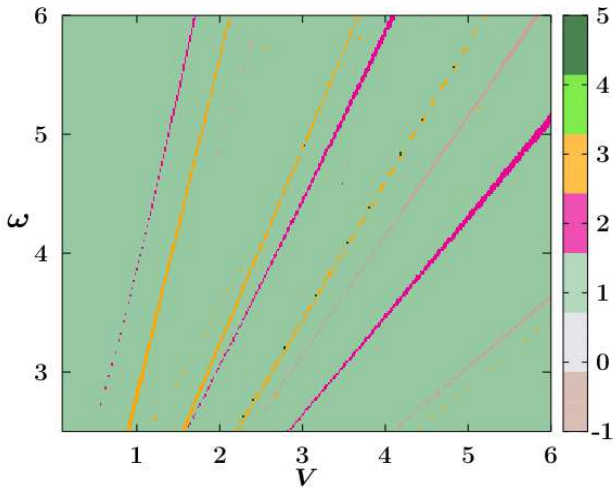}
\caption{\label{Fig Phase diagram II high-freq} A similar phase diagram as the previous figure is presented in the high-frequency regime $\omega > 2$. Here, {\it nine} photon sectors are considered and the parameters are fixed at: $\phi = -\pi/4$, $g_A = g_B = 0.1$ and $h = 0.2$. The color bar represents the Chern number $C_{\rm s}$.}
\end{figure}

For the lower frequency regime, the phase diagram is shown in Fig. \ref{Fig Phase diagram II low-freq}. This phase diagram displays the sum of the Chern numbers $C_{\rm s}$ of all the Floquet bands below $E=0$. The Chern number $C_{\rm s}$ is playing the role of the topological invariant. Here we consider the driving amplitude in the range $0.1\leq V\leq 6.0$. In this regime, around the left-bottom corner of the plot marked by a white region, we observe huge values of $|C_{\rm s}|$. We have marked the different regions of the phase diagram by colors only for $|C_{\rm s}| \leq 10$. For smaller values of both $\omega$ and $V$, we see many jumps in the values of $C_{\rm s}$ which indicate rapid Floquet topological phase transition.  

A major portion of this phase diagram is occupied by the topologically non-trivial phase with $C_{\rm s} = 1$ (light-green color). We have found some scattered regions where $|C_{\rm s}| >1$. However, among these different higher Chern number regions, we find that the regions with $|C_{\rm s}| = 2,\, 3,\, 5$, and $9$ are prominent. The other values of the Chern number are observed as many tiny islands in the phase diagram, which are mostly located near the $V$-axis with $V \gtrsim 2$. Since here the frequency is low enough, we have to consider {\it eleven} photon sectors to achieve convergence in our calculation. A qualitatively similar topological property was reported for the case of periodically driven honeycomb lattice \cite{mikami2016brillouin}. However, here we observe this in a 1D system.

In Fig. \ref{Fig Phase diagram II high-freq}, we have presented the phase diagram in the higher frequency regime $(2.0< \omega \leq 6.0)$. Here we consider the same range of driving amplitude $V$ as earlier. Because of the high-frequency regime, we need to consider only {\it five} photon sectors to get convergence. However, here we present results with {\it nine} photon sectors. Here again, we mostly observe a ``sea" of $C_{\rm s}=+1$ region, but the regions with higher values of $C_{\rm s}$ are much smaller than the low-frequency case. We can see some strips of regions with $C_{\rm s} = 2$ and $3$. Here, we observe islands of minuscule sizes with the Chern number $C_{\rm s} = 5$, which is the highest observed value of $C_{\rm s}$ for this high-frequency regime.

The Floquet topological phase transition discussed in this subsection can be understood by investigating the Berry curvature, whose integration over the BZ gives the Chern number. We know that when a (Floquet) topological phase transition takes place, the Berry curvature localizes near the band touching points in the BZ. In Fig. \ref{Fig Phase diagram II high-freq}, we see tiny dark-green islands with $C_{\rm s} = 5$ on the orange strip with $C_{\rm s} = 3$. These points are the critical points of the Floquet topological phase transitions $C_{\rm s} = 3 \leftrightarrow C_{\rm s} = 5$. Therefore, in the next subsection, we study in detail the cases of higher Chern numbers $C_{\rm s} = 3$ and $5$ at the high-frequency regime. At this regime, the interactions among different photon sectors are minimal. As a consequence, we have found that the summed over Chern number $C_{\rm s}$ is equal to the Chern number $C$ of the lower Floquet band of the zero photon sector. This happens because each photon sector with $m <0$ contributes a pair of Chern numbers with the same magnitude but of opposite signs. Consequently, when we sum them up, they cancel each other, and only the Chern number corresponding to the lower Floquet band of the zero photon sector survives, and thus it leads to $C_{\rm s} = C$. 

\subsection{\label{Higher topology case: $C=3$} Cases of the higher Chern numbers at the high frequency limit: $\bs{C=3}$ and $\bs{5}$}

\begin{figure}[b]
\includegraphics[width=8.5cm,height=2.8cm]{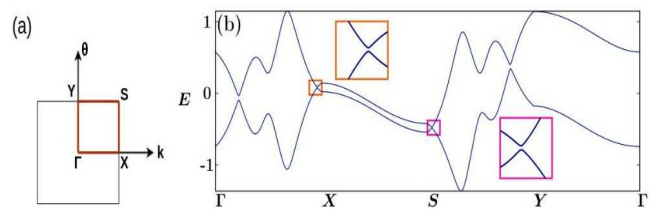}
\caption{\label{Fig band diagram} The Floquet bands are presented for $C=3$ case along the high-symmetry lines. Here we set $V=1.0$ and $\omega = 2.8$. Subfigure (a) demonstrates the high symmetry path $\Gamma-X-S-Y-\Gamma$ in the BZ. Here, in the $(k, \theta)$ coordinates, the symmetry points are at $\Gamma = (0, \pi),\,X = (\pi, \pi),\,S = (\pi, 2\pi),$ and $Y = (0, 2\pi)$. Subfigure (b) shows that there are many regions where Floquet bandgaps are very small. This behavior is responsible for the localization in the Berry curvature and consequently its leads to high Chern number.}
\end{figure}

\begin{figure}[t]
\includegraphics[width=9cm,height=4cm]{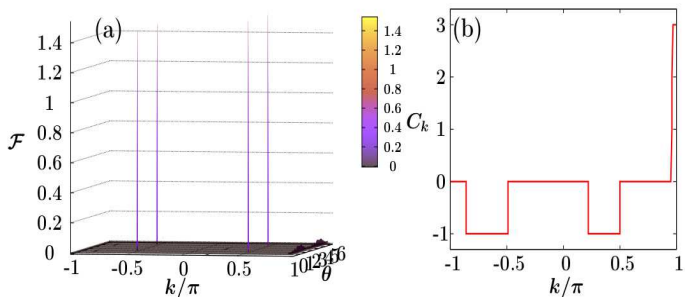}
\caption{\label{Fig C3 Berry curvature} For $C=3$, the Berry curvature is plotted in (a) as a function of $k$ and $\theta$ for the lower Floquet band of the zero photon sector. The driving parameters are fixed at $V=1.0$ and $\omega = 2.8$; and here {\it nine} photon sectors are considered. Fixing the parameters same as (a), $C_k$ is plotted as a function of $k$ in (b) to observe the contribution from the different parts of the BZ in the Chern number.}
\end{figure}

\begin{figure}[b]
\includegraphics[width=8.5cm,height=7cm]{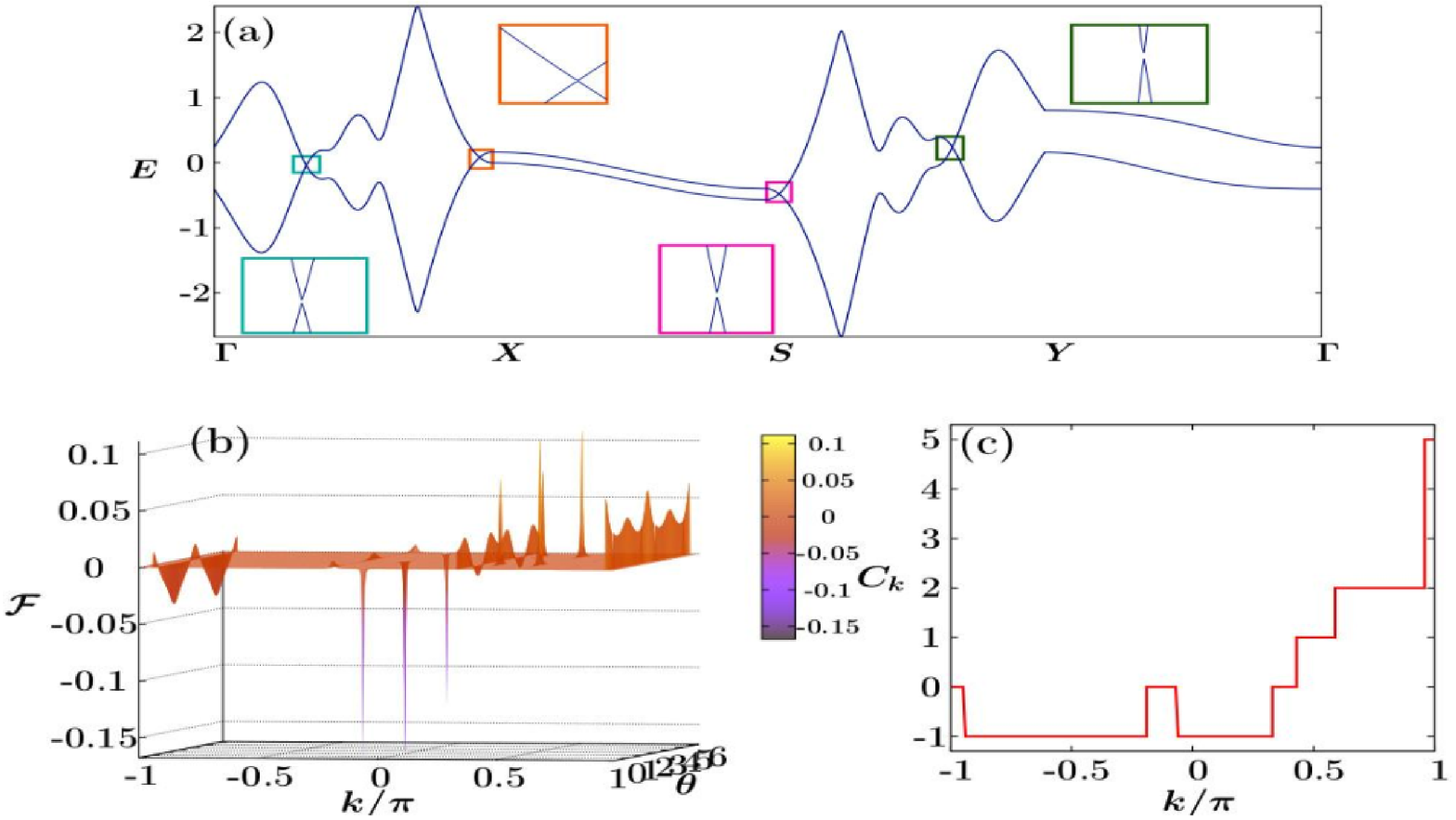}
\caption{\label{Fig C5 Berry curvature} The driving parameters are set at $V=3.02$ and $\omega = 4.9$; and {\it nine} photon sectors are considered for the numerics. Here for the lower Floquet band the Chern number $C=5$. Subfigure (a) presents the Floquet band diagram along the same high symmetry path $\Gamma-X-S-Y-\Gamma$. Here we also find multiple points where the bandgaps are tiny. Again along $X-S$ path, the behavior of the Floquet bands is similar to the above case of $C=3$. In (b), the Berry curvature is plotted as a function of $k$ and $\theta$ for the lower Floquet band of the zero photon sector. Subfigure (c) presents $C_k$ as a function of $k$.}
\end{figure}

We now focus on two cases of higher Chern numbers with $C=3$ and $5$. Here we consider the regime of high-frequencies in $\omega$ and strong driving strengths $V$. 

\subsubsection{Case: $\bs{C=3}$}

In Fig. \ref{Fig band diagram}(a), we have shown a high symmetry path $\Gamma-X-S-Y-\Gamma$ by solid lines in the BZ. For a case of $C=3$, we have then demonstrated in Fig. \ref{Fig band diagram}(b) the Floquet bands corresponding to $m=0$ photon sector along the high symmetry path. Here we see a couple of points surrounded by the rectangles, where the Floquet bands are almost closing with tiny finite bandgaps. We have zoomed those regions to show clearly the presence of the bandgaps. The presence of gaps is allowed us to calculate the Chern number of the Floquet bands. If we tune the driving parameters $(V, \omega)$ slightly, we see band closing and reopening with a different Chern number ($C=1$). Another interesting region is observed when we consider the high-symmetry path $X-S$. Here two bands are always nearby and varying almost parallel as we change the parameter $(\theta: \pi \rightarrow 2\pi)$ along the $X-S$ path.   

It is well known that the bandgap closing and opening are responsible for the localization of the Berry curvature in the reciprocal space. We have seen above that the bandgaps between the Floquet bands in the zero photon sectors are small at different points within the BZ. Therefore, we now investigate the effect of this behavior of the Floquet bands on the Berry curvature and its consequences in the Chern number. The results are shown in Fig. \ref{Fig C3 Berry curvature}. According to our expectation, we see in Fig. \ref{Fig C3 Berry curvature}(a) localization in the Berry curvature around the regions where the bandgaps are very small. Since the Chern number is obtained by integrating the Berry curvature over the BZ, we investigate the contribution of the Berry curvature from different regions of the BZ to the Chern number. For this purpose, we define a measure {\it local Chern number} $C_k$ as
\begin{equation}
C_k = \frac{1}{2\pi}\int\limits_{k^\prime = -\pi}^{k}\!\!\!\! dk^\prime \!\!\int\limits_{\theta = 0}^{2\pi} \!\! d\theta\,\, \mathcal{F}(k^\prime,\theta). 
\label{eq: ChernNumber wrt k}
\end{equation}
According to this definition, the Chern number and the local Chern number are related as $C = C_\pi$. In Fig. \ref{Fig C3 Berry curvature}(b), we show how $C_k$ changes with $k$. Here we observe transitions in $C_k$ around those regions where the Floquet bandgaps are small. This is happening because the Berry curvature is nonzero exactly around those regions. Interestingly, except when $k \simeq \pi$, the local Chern number $C_k$ is transitioning only within $0 \leftrightarrow (-1)$. These are happening exactly at those regions where the Floquet bandgaps are very small, as well as the bands are linearly dispersive. The transition $0 \rightarrow 3$ at $k \simeq \pi$ can be understood by comparing the properties of the Floquet bands along the high-symmetry path $X-S$, which is defined as $\pi \leq \theta \leq 2\pi$ and $k = \pi$. Here, two Floquet bands vary in parallel, maintaining almost a fixed but small bandgap. In Fig. \ref{Fig C3 Berry curvature}(a), we see the localization of the Berry curvature exactly along a line that is the same as the $X-S$ path.

\subsubsection{Case: $\bs{C=5}$}

The results for this case are shown in Fig. \ref{Fig C5 Berry curvature}. In Fig. \ref{Fig C5 Berry curvature}(a), the Floquet band diagram is presented along the same high symmetry path $\Gamma-X-S-Y-\Gamma$ for the driving parameters $V=3.02$ and $\omega = 4.8$. Here again, we observe multiple points where the Floquet bandgaps are very small with linear dispersive bands. Moreover, along the $X-S$ path, the qualitative behavior of the Floquet bands is similar to the case of $C=3$. We present the Berry curvature in Fig. \ref{Fig C5 Berry curvature}(b). This figure shows the correlation between the localization of the Berry curvature and the Floquet bandgaps. According to our expectation, here again, we see that the smaller Floquet bandgaps are responsible for the stronger localization and {\it vice versa}. Like the previous case, we also see in Fig. \ref{Fig C5 Berry curvature}(c) the appearance of nonzero Berry curvatures at different regions in the BZ, and that leads to the transitions in $C_k$. From the transition perspective, the property of the local Chern number is very rich here. We observe many different transitions in $C_k$. This is clearly the consequence of the complex structure of the Berry curvature as shown in Fig. \ref{Fig C5 Berry curvature}(b). In this case, besides the transitions of $0 \leftrightarrow (-1)$, we also observe transitions like $0 \rightarrow 1$, $1 \rightarrow 2$, and $2 \rightarrow 5$. The final transition of $2 \rightarrow 5$ is happening again due to the same behavior of the Floquet bands along the high symmetry path $X-S$ as discussed above for the case of $C=3$. In case of $C=3$, the final transition of $C_k$ was $0 \rightarrow 3$. The equal amount of jump in local Chern number $\Delta C_k = 3$ around $k=\pi$ suggests that the contribution of the $X-S$ path to the Chern number $C$ is exactly the same for both the high Chern number cases.  

\section{\label{sec:Final remarks} Summary and final remarks}

In this paper, we study a periodically driven E-SSH model. An extensive study of different phase diagrams clarifies that the external periodic driving makes this system topologically much richer. First, we have found that our periodic driving protocol introduces gaps between the Floquet bands at the BZ boundary for non-zero values of the parameter $\phi$. In the case of the undriven E-SSH model, the parameter $\phi$ determines the topology of the system. This result then motivates us to study a phase diagram in the parameter space of the relative strength of the NNN hopping and the parameter $\phi$. We have observed that, for the weaker driving field strength, the phase diagram is almost identical ($C_{\rm{s}}=0\, {\rm or} \pm1$) to the phase diagram corresponding to the undriven system except that the sign of the Chern number is now opposite but of the same magnitude. However, for the stronger driving strength, the Chern number not only changes its sign, but its magnitude also increases with $C_{\rm{s}}= \pm3$. Moreover, we observe that the larger Chern number regions are surrounded by lower Chern number regions with $C_{\rm{s}}=\pm1$. The discovery of higher Chern numbers for the stronger driving field motivates us to extensively study the role of the driving parameters ($\omega$ and $V$) on the topological properties of the system. 

We then study the topological phase diagram by studying the Chern number $C_{\rm{s}}$ as a function of the driving frequency $\omega$ and driving amplitude $V$. Here we concentrate separately on the driving frequency in two regimes: (1) lower than or the same order of the bandgap of the undriven E-SSH model; and (2) higher than the bandgap of the undriven E-SSH system. In the lower frequency regime, we observe many sharps topological phase transitions. Particularly for the weaker driving amplitudes, we observe topological phase transitions between a non-trivial phase with a very high Chern number to another non-trivial phase with another high Chern number. Higher Chern number is also observed in the high-frequency regime but within a very restricted value of the driving parameters. Moreover, in this regime, we have observed the highest value of the Chern number as $C_s = +5$. The above discussion clearly shows that the 1D E-SSH model under monochromatic periodic driving can have many different topological phases with a much higher Chern number depending on the driving parameters.

We investigate in detail the higher Chern number cases of $C=3$ and $5$ at the high-frequency regime. Here we observe that whenever two Floquet bands in the zero photon sector come close to each other, the Berry curvature localizes there, and consequently, it contributes to the Chern number of the Floquet bands. In order to identify the contribution of the Berry curvature to the Chern number from different regions of the BZ, we introduce a measure ``local Chern number" $C_k$. We see many jumps in the values of $C_k$ when we plot this as a function of $k$, and these jumps exactly occur at those values of $k$ where the gaps in two Floquet bands are very small. 

In this study we consider a driving protocol of the form $V(t) = V_1 \cos \omega t + V_2 \sin \omega t$. Following the recently proposed perturbation theory in Ref. \cite{Goldman-Dalibard}, this smooth driving protocol can be realized at the high-frequency limit (very large $\omega$) by a class of four-step pulse sequences of the form 
\[\{H_0 + V_1, H_0 + V_2, H_0 - V_1, H_0 - V_2\},\] where $H_0$ is the undriven Hamiltonian. Here each pulse is applied for the time interval $T/4$, where $T = 2\pi/\omega$. This kind of pulse sequences can be realized in cold atoms and optical lattice setup \cite{Optical-Lattice01,Optical-Lattice02,Optical-Lattice03}.   

\begin{acknowledgments}
Authors acknowledge financial support from DST-SERB, India through the Core Research Grant CRG/2020/001701.
\end{acknowledgments}

\bibliography{References} 

\end{document}